# Multiscale Dynamics of Lipid Vesicles in Polymeric Microenvironment

Selcan Karaz [1], Mertcan Han [2], Gizem Akay [3], Asim Onal [4], Sedat Nizamoglu [2,3,4], Seda Kizilel [1,4] and Erkan Senses [1,5]

[1] Department of Chemical and Biological Engineering, Koc University, Sariyer, Istanbul 34450, Turkey;

[2] Department of Electrical and Electronics Engineering, Koc University, Istanbul 34450, Turkey

[3] Department of Materials Science and Engineering, Koc University, Sariyer, Istanbul 34450, Turkey;

[4] Graduate School of Biomedical Science and Engineering, Koc University, Istanbul 34450, Turkey;

[5] Koç University Surface Science and Technology Center (KUYTAM), Rumelifeneri Yolu, Sariyer, Istanbul 34450, Turkey

**Abstract**

Understanding dynamic and complex interaction of biological membranes with extracellular matrices plays a crucial role in controlling a variety of cell behavior and functions, from cell adhesion and growth to signaling and differentiation. Tremendous interest in tissue engineering has made it possible to design polymeric scaffolds mimicking the topology and mechanical properties of the native extracellular microenvironment; however, a fundamental question remains unanswered: that is, how the viscoelastic extracellular environment modifies the hierarchical dynamics of lipid membranes. In this work, we used aqueous solutions of poly(ethylene glycol) (PEG) with different molecular weights to mimic the viscous medium of cells and nearly monodisperse unilamellar DMPC/DMPG liposomes as a membrane model. Using small-angle X-ray scattering (SAXS), dynamic light scattering, temperature-modulated differential scanning calorimetry, bulk rheology, and fluorescence lifetime spectroscopy, we investigated the structural phase map and multiscale dynamics of the liposome–polymer mixtures. The results suggest an unprecedented dynamic coupling between polymer chains and phospholipid bilayers at different length/time scales. The microviscosity of the lipid bilayers is directly influenced by the relaxation of the whole chain, resulting in accelerated dynamics of lipids within the bilayers in the case of short chains compared to the polymer-free liposome case. At the macroscopic level, the gel-to-fluid transition of the bilayers results in a remarkable thermal-stiffening behavior of polymer–liposome solutions that can be modified by the concentration of the liposomes and the polymer chain length.

# 1. Introduction

Liposomes have been widely studied as models of real cell membranes to understand the complex relationship of the cells with their environment and to understand membrane dynamics at macroscopic and molecular levels [1]. They are formed by self-assembling hydrophilic head groups and hydrophobic acyl tails to spontaneously form closed vesicles in aqueous media and reflect the similarity of the protein-free parts of the real membranes [2]. The interaction of phospholipids is not controlled by covalent bonds; instead, the membrane lipids are held together by weaker forces, such as van der Waals forces and hydrophobic interactions [3], making the membranes susceptible to the physical and chemical changes in their microenvironment and adjust their fluidity, stiffness, and thickness accordingly [4,5]. Polyethylene glycol (PEG) is a commonly used synthetic polymer to mimic the crowded macromolecular medium and ECM due to its non-toxic structure, tunable properties, and flexible conformation [6]. Liposome-PEG complexes have therefore found unique applications in drug delivery and tissue engineering due to the similarity of liposomes to real cell membranes and the steric protective effect of linear PEG chains [7–9]. The lipid nanoparticles used in mRNA-based COVID-19 vaccines are recent examples of these composite systems. PEG is used to functionalize the surface of the liposomes to enhance their colloidal stability and biodistribution [10,11]. Despite controversial evidence, conjugated PEG on liposomes has been posited as the reason for rare allergic responses to these vaccines, and this allergen effect was shown to be highly dependent on molecular weight [12]. Additionally, as the local viscoelasticity depends strongly on the length of the attached linear chains, the lipid nanoparticles can essentially experience a very different microenvironment compared to their bulk solution, which can alter the delivery efficiency of vaccines and other drug molecules. Furthermore, because the segmental and collective chain relaxation of the polymers occur at timescales comparable to those of the single lipid motion and membrane undulations, the dynamics of the flexible polymer chains and the liposomes are expected to be coupled [13]. Despite the practical use of polymer–lipid nanocarriers, a thorough understanding of the effect of the polymeric microenvironment on membrane dynamics is still lacking.

It is known that lipid bilayers can form intermolecular hydrogen bonds and bind to the specific sites of proteins that can enable long-distance intraprotein and protein–lipid coupling [14–16]. Similarly, it was shown that PEG favors hydrogen bonding due to the hydrophilic head groups of the lipids and ether oxygen of PEG [17,18]. According to the MD simulation results, the penetration of PEG inside the membrane is not observed. Instead, it alters the membrane structure in such a way that it occurs at the bilayer/water interface [19]. The attractive interaction may result in a formation of a stable polymer-bound layer on liposomes (which is strongly dependent on molecular weight) [18], yet the effects of the polymer layer on the membrane structure and dynamics have not been elucidated, although some macroscopic effects have been reported. For example, PEG, a highly hydrated polymer, can disrupt packing in the membrane monolayers, which allows lipid fusion to be induced [20,21]. Additionally, with the existence of PEG outside of the membranes, cell damage resulting from hypothermia can be prevented. In one study, PEGs with lower molecular weights were found to destabilize the lipid monolayer due to high osmotic pressure exerted on the membrane, although PEG 35,000 Da showed stronger stabilization, with successful applications on cold-induced injuries during organ transplantation [22]. PEG can also alter bilayer curvature and lateral packing of membranes; in turn, it causes some modulations in thermal behavior [23]. Phase transition of DMPC vesicles with PEG molecular weights ranging from 200 to 20,000 Da and varying concentrations were investigated in [24]. It was shown that the effect of PEG on the phase transition temperature of the lipids is dependent on molecular weight and concentration. In another study on the interaction of multilamellar DMPC, DPPC, and DSPC liposomes with PEG molecules, it was suggested that the main phase transition temperature is affected more rapidly compared to the pretransition temperature with increased PEG concentration [25]. Dehydration of lipid membranes is also known to be dependent on molecular weight [26]. These results from separate works suggest that PEG can strongly alter the structural characteristics of the liposome, although a comprehensive physicochemical study is needed for complete understanding.

With respect to dynamics, the viscosity of the lipid bilayers is related to membrane fluidity and permeability, which affect cellular processes contributing to many critical activities [27,28]. Lipid motion occurs over a broad range of length and time scales [4,5,29–31]. For example, lipid flip-flop motion takes place at large time scales and has an impact on maintaining the composition of lipids on the inner and outer sides of the membrane [32], whereas cell signal transduction is affected at the molecular length scale by lateral diffusion of lipids within the membrane [33]. On the other hand, the collective motion of lipids within the membrane results in coherent movements, such as bending, stretching, and thickness fluctuations, which are responsible for pore and peptide channel formation [34] and are crucial for cytokinesis [35], motility, and extracellular vesicle formation [36]. Microviscosity of the lipid bilayer is closely related to the macroscopic transport properties of the membrane and has been measured with various fluorescence techniques, including time-resolved fluorescence spectroscopy, fluorescence anisotropy, and determination of fluorescence quantum yield [37–39]. The microviscosity of DMPC membranes was estimated to change from 1.5 cP to 97 cP after phase transition to the gel state [40]. At the macroscopic level, the diffusion of liposomes is closely related to the bulk viscosity of the surrounding medium. All these studies strongly suggest that lipid membrane interaction with the surrounding polymeric media can alter the membrane structure and dynamics. Although liposome–polymer complexes have been widely utilized in drug delivery and tissue engineering applications in the past [41,42], a fundamental question remains unanswered; that is, how the polymeric medium affects the bulk viscous properties and microviscosity of membranes.

In this work, we used aqueous solutions of poly(ethylene glycol) (PEG) with a wide range of molecular weights (from 1.5 kDa to 400 kDa) to mimic the viscous medium of the cells and nearly monodisperse unilamellar vesicles of DMPC/DMPG liposomes to model cell membranes. Using small-angle X-ray scattering (SAXS), dynamic light scattering (DLS), temperature-modulated differential scanning calorimetry (DSC), bulk rheology, and fluorescence lifetime spectroscopy (FLS), we investigated the structural phase map and dynamics of the liposome–polymer mixtures from the nanoscale to the bulk scale. The results show that the microviscosity of the lipid bilayers is strongly influenced by the relaxation of the whole chain, resulting in accelerated dynamics of the lipids within the membrane (even in the gel state) compared to the case of polymer-free liposomes. At the macroscopic level, the gel-to-fluid transition of the bilayers causes an unprecedented thermal-thickening behavior of the polymer–liposome solutions, which is controlled by the multilamellarity, size, and concentration of the liposomes.

## 2. Materials and Methods

### 2.1. Preparation of Liposome Solutions

Lipids (1,2-dimyristoyl-sn-glycero-3-phospho-choline (DMPC) and 1,2-dimyristoyl-sn-glycero-3-phospho-(1'-rac-glycerol) (sodium salt) (DMPG)) were purchased from Avanti Polar Lipids (Alabaster, AL, USA). Lipids containing (5% *w/w*) DMPG and DMPC were dissolved in chloroform and mixed for one hour. The solution was then evaporated in a vial, and the thin film of the lipids was kept under a vacuum at 80 °C for 24 h. The film was hydrated with DI water to obtain 20 mg/mL of liposome suspension and vortexed for 2 h for complete dissolution. Two different lipids were used because DMPG is a charged phospholipid, increasing the stability and the dispersion of the system. Lipid suspensions were extruded using a mini extruder (Avanti Polar Lipids, Alabaster, AL, USA) after 15 min of temperature equilibration. By using a heating block, the temperature of the extruder was kept at 40 °C at all times. To prepare the unilamellar vesicles, the lipid suspensions were subsequently extruded 15 times through 400 nm membranes, 15 times through 200 nm membranes, and 41 times through 100 nm membranes. Extruded liposome suspensions were used immediately after preparation or stored at 40 °C while mixing in a vortex. All solutions were used within 24 h. The stability of the liposomes was confirmed using dynamic light scattering (DLS) before the experiments.

*2.2. Preparation of Liposome Solutions Containing Oil Red O*

Oil Red O was purchased from Sigma-Aldrich and used as received. For fluorescence lifetime spectroscopy experiments, $2 \times 10^{-5}$ uM Oil Red O was added to the same amount of lipids, and the liposome suspensions were prepared using the same protocol described above.

*2.3. Preparation of Liposome–Polymer Mixtures*

Poly(ethylene glycol) (PEG) with four different molecular weights (1.5 kDa, 20 kDa, 100 kDa, and 400 kDa) was purchased from Sigma-Aldrich and used as received. Stock solutions of polymers with M < 400 kDa were prepared with a concentration of 100 mg/mL. PEG 400 kDa was prepared at 50 mg/mL due to its high viscosity. Required amounts of the neat polymer solutions were added to the liposome solution slowly, and the final concentration of polymers was kept constant at 20 mg/mL. The liposome–polymer mixture was placed in a vortex for 15 min for equilibration prior to use.

*2.4. Dynamic Light Scattering (DLS)*

Hydrodynamic size, size distributions, and diffusion constants of liposomes were determined by dynamic light scattering (DLS, Malvern ZS Zetasizer) with backscatter detection at a scattering angle of 173° and a wavelength of 633 nm. Samples were diluted to 0.2% concentration for measurement. In order to investigate the effect of temperature on liposome size, samples were measured by changing the temperature in a stepwise manner from 10 °C to 40 °C at a rate of 2 °C/min with 120 s equilibration time for each step. Each measurement was repeated three times, and the average values are reported.

*2.5. Differential Scanning Calorimetry (DSC)*

The phase transition temperatures of the liposomes from fluid to gel phase were measured using TA Instrument's Q25 AutoDSC. Next, 10–15 mg of samples was sealed in Tzero pans with hermetic lids. An empty pan was measured as reference. Samples were first heated to 60 °C and equilibrated for 5 min. Then, they were cooled down to 5 °C at a rate of 2 °C/min. The crystallization temperatures of liposomes were obtained using the cooling curve for each case.

Enthalpy values were calculated from the area of exothermic peaks. The cycles were repeated several times to ensure reproducibility of the data.

*2.6. Rheology*

Viscosity measurements of the solutions were performed using an Anton-Paar rheometer (MCR-302). A conic plate with a 50 mm diameter and a cone angle of 1° was used (CP50-1). Temperature was controlled using a Peltier control device coupled with a hood to prevent evaporation. The temperature of the system was continuously changed from 10 °C to 40 °C with a heating rate of 1 °C/min. Data were collected every 60 s at a constant shear rate of 30 $s^{-1}$.

To obtain the calibration curve of microviscosity calculations, the viscosity of the glycerol–methanol–Oil Red O solutions with varying ratios was measured by a conic plate with a 50 mm diameter and a cone angle of 2° (CP50-2) at 15 °C and 35 °C. The shear rate was changed from 1000 $s^{-1}$ to 1 $s^{-1}$, and the viscosity data were collected at a shear rate of 10 $s^{-1}$ for all solutions.

*2.7. Small-Angle X-Ray Scattering (SAXS)*

SAXS data were collected using an Anton Paar SAXSpoint 5.0 with a sample-to-detector distance of 500 mm. The samples were sealed in a quartz capillary holder and placed on a heated/cooled sampler for temperature control. Frames were collected for 3 min periods over a total duration of 1.5 h per sample. The transmission data were also obtained for each sample and the background (water only) for proper reduction and subtraction. Data reduction was performed using the SAXS analysis program of the SAXSPoint 5.0.

*2.8. Cryo-TEM*

Each sample was placed on a 400 mesh copper grid and transferred to a Gatan cryo transfer holder containing liquid $N_2$. The samples were then investigated by cryogenic transmission electron microscopy (cryo-TEM) at 120 kV (Hitachi HT7800, Tokyo, Japan).

*2.9. UV-Vis Spectroscopy*

UV-vis spectra were collected in the wavelength range of 400 to 700 nm with a Shimadzu (Kyoto, Japan) UV-3600 -UV-VIS-NIR spectrophotometer in transmission mode using quartz cuvettes at 35 °C for 5 min for equilibration before each measurement. DI water was used as a reference of neat liposome solutions. For liposome and PEG mixtures, neat PEG solutions with appropriate molecular weights were used as reference.

*2.10. Time-Resolved Photoluminescence*

Time-resolved spectroscopy measurements were carried out with a Picoquant MicroTime 100 (PicoQuant GmbH, Berlin, Germany) time-resolved confocal fluorescence microscope. The picosecond diode laser ($\lambda_{ex}$ = 375 nm) pulsed at 10 MHz repetition rate with 8 mW power was used as the excitation source via an Olympus PlanC N 40×/0.65, FN 22, microscope objective. A single-photon sensitive detector (PMA Hybrid 50) based on a photomultiplier tube (R10467 from Hamamatsu) and a time-correlated single photon counting system (HydraHarp 400) were utilized. The data were analyzed in fluorescence lifetime imaging (FLIM) mode with SymPhoTime 64 (Picoquant) software. PL decays were fit by a two-exponential decay, and the average lifetime was calculated from an intensity-weighted mean. Each measurement was repeated three times.

Throughout this manuscript, the error bars represent one standard deviation.

## 3. Results and Discussion

*3.1. Structure*

We first investigated the structure of the liposomes in their neat form and in polymer solutions. We prepared neat (polymer-free) DMPC/DMPG liposome solutions using a common extrusion method (see the experimental section for details). The size and polydispersity of the resulting liposomes were determined using DLS (see Figure S1 and Table S1). The unextruded solution contained highly polydisperse micron-sized liposomes, whereas the extrusion process progressively decreased the size of the liposomes to 119.9 ± 2.54 nm, with a polydispersity index of 0.075 ± 0.010, yielding a monodispersed vesicle solution. The morphology of a representative liposome was also visualized by cryo-TEM (shown in Figure 1).

We evaluated the lamellarity of the neat vesicles using DSC and SAXS. Phospholipids do not undergo a usual phase transition from solid to liquid form with temperature; instead, they make conformational changes. Figure 1B shows the specific heat flow curves (cooling) obtained for the 20 mg/mL liposome solutions (nominal lipids) in different stages of the extrusion process. The hydrophobic tails of DMPC and DMPG are identical, and their transition temperature from gel to fluid state is observed at ≈24 °C [43]. The sharp and intense peak at ≈22.4 °C for the unextruded liposome solution occurs due to a large fraction of multilamellar vesicles that contain highly ordered lamellar structures in the gel state. The intensity decreases gradually as the size of the liposomes is decreased, suggesting a transition from a multilamellar to unilamellar structure upon extrusion. In the final 100 nm vesicle solution, the peak completely disappears; only a broad transition at ≈23 °C is observed. Below the melting temperature, lipids present with a closely-packed and ordered structure, whereas the long-range order in acyl tails is disrupted in the ripple phase during transition and completely disappears in the fluid phase at higher temperatures [44]. The differences in the interbilayer interactions result in unilamellar vesicles with a higher degree curvature, lower lateral packing, and different interchain interactions, which account for the broadened transition observed for the unilamellar structures, in agreement with the previous studies [45–47]. Nevertheless, the DSC curves around the transition temperature suggest that ≈100 nm monodisperse vesicles are primarily unilamellar.

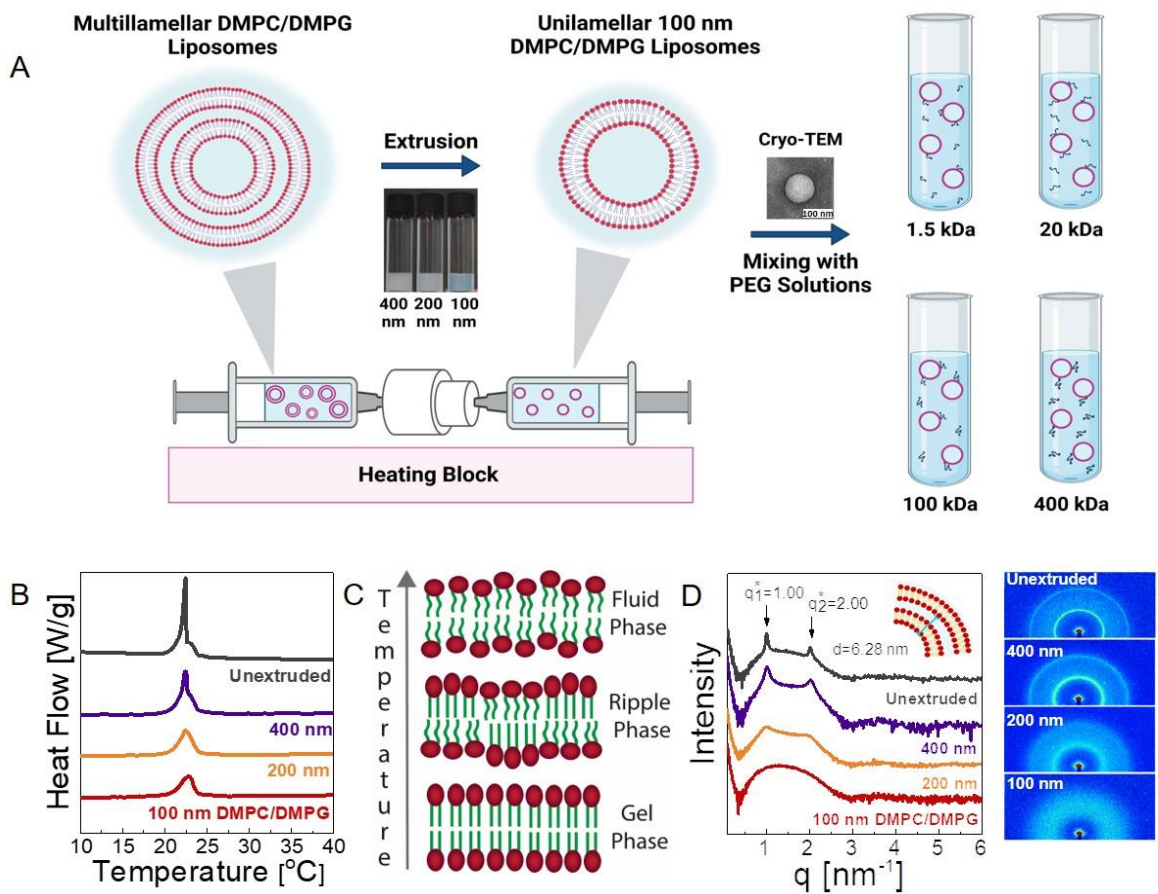

**Figure 1.** (**A**) Schematic image of the system; 20 mg/mL DMPC/DMPG solution is extruded through 400 nm (15×), 200 nm (15×), and 100 nm (41×) membranes, respectively. Multilamellar liposomes forming spontaneously in solution are represented on the left-hand side before extrusion. Following extrusion, unilamellar and 100 nm liposomes are obtained. The liposome solutions are then mixed with solutions of PEG with four different molecular weights, specifically 1.5, 20, 100, and 400 kDa. (**B**) DSC traces of 20 mg/mL unextruded, 400 nm, 200 nm, and 100 nm lipid solutions. (**C**) Phase transition of lipids from gel phase to fluid phase with increasing temperature. (**D**) SAXS analysis of unextruded, extruded 400 nm, 200 nm, and 100 nm liposomes and raw detector data. Two peaks at $q_1^* = 1.00$ and $q_2^* = 2.00$, corresponding to Bragg spacing of 6.28 nm, were obtained. All measurements in B and D were repeated three times to ensure reproducibility of the data.

SAXS offers more direct evidence of lamellarity (see the experimental section for details). Figure 1D shows scattering intensity as a function of wave vector, q, for unextruded, 400 nm, 200 nm, and 100 nm DMPC/DMPG solutions (see also the raw detector images in Figure 1D). A characteristic set of two equally spaced peaks was observed for unextruded multilamellar liposomes at $q_1^* = 1.0$ nm$^{-1}$ and $q_2^* = 2.0$ nm$^{-1}$, corresponding to the Bragg spacing of $d = 2\pi/q_1^* = 6.28$ nm. This is equal to the repeat distance of the stacked lipid bilayers, including water in between (see the sketch in the inset of Figure 1D) and in agreement with the previously reported values for multilamellar vesicles [48]. The intensity of the multilamellar peaks decreased gradually (without changing their *q* positions) as the size of the liposomes decreased by extrusion. For 100 nm liposomes, the peaks completely disappear, and the liposomes become exclusively unilamellar. Overall, DLS, DSC, and SAXS results conclude that the 100 nm diameter polymer-free liposomes are nearly monodisperse and unilamellar, whereas polydisperse 200 and 400 nm liposomes contain a significant amount of multilamellar structures. Because the unilamellar vesicles with well-controlled permeability, viscosity, and polarity are representative of the protein-free part of the lipid bilayer of biological cells, our analysis of the polymer–liposome mixtures in the following sections is be mostly based on 100 nm unilamellar liposomes.

For polymer–liposome solutions, we used different molecular weights of linear PEG, specifically 1.5, 20, 100, and 400 kDa. Previous studies showed that the thermal transition of the lipid membranes can be dependent on molecular weight [24,49], which implies that the structural dynamics of the membranes can change with the length and size of the polymer chains. Additionally, the molecular weight of the polymers controls other key properties, such as osmotic pressure and viscosity. Therefore, a wide range of PEG molecular weights was chosen, which resulted in hydrodynamic radii ranging from 0.78 ± 0.10 nm to 18.17 ± 3.28 nm (see Figure S2 for DLS data of the neat polymers) as the molar mass of PEG changed from 1.5 kDa to 400 kDa. Hydrodynamic sizes obtained by DLS are close to the theoretical radius of gyration ($R_g$) values of PEGs (tabulated in Table S2), suggesting that the polymers do not form clusters in water [50].

We then prepared polymer–liposome solutions containing 1.7 wt% lipids and 1.7 wt% polymer in water. The stability of the liposomes in the polymer solutions were investigated through DLS using dilute solutions (see Figure S2A) showing the hydrodynamic diameters of 100 nm DMPC/DMPG liposomes in solutions with different polymer molecular weights. A representative cryo-TEM image of the composite solution is also presented in Figure S2B . The size and polydispersity index (PDI) values are presented in Table 1. The diameter of the liposomes gradually increases with increasing molecular weights in comparison with the neat 100 nm vesicles, suggesting that there is an attractive interaction between liposomes and PEG chains, resulting in an adsorbed polymer layer on bilayers (see Figure 2B). Membrane lipids tend to interact with their environment via weaker forces, such as van der Waals and electrostatic interactions [51]. The interaction between PEG and lipids was shown to be hydrogen bonding between the ether oxygen of PEG and hydrophilic head groups. The binding of PEG onto PC lipid bilayers was recently demonstrated by Zhang et al. [18],where the bound fraction is enhanced in the presence of hydroxyl-containing PC lipids. Moreover, as PEG has terminal $OH^-$ end groups, which make it hydrophilic, it prefers to stay in the bilayer/water interface due to energetic barrier, as shown by MD simulations [19]. On the other hand, the penetration of PEG into membranes is possible if the terminal end groups of polymers are modified to a relatively more hydrophobic group. De Mel et al. showed that the alkyl group of n-Alkyl-PEO can reside within the bilayer structure, whereas the unmodified hydrophilic part of the polymer prefers to stay on the outside of the membrane, which is similar to the case of our study; thus, the polymer forms a bound layer on the outer surface of the liposomes, as suggested by DLS [52]. Comparison of the size increase in the liposomes in PEG solutions with respect to neat 100 nm liposomes allows for estimation of the hydrodynamic thickness of the adsorbed layer, defined as $\delta$. Figure 2C shows the dependence of $\delta$ on the PEG $M_w$, and its value ranges from ≈1 nm to ≈20 nm. The values are in close agreement with the theoretical Rg values of PEG at respective molecular weights under theta conditions. Moreover, a linear relationship with $\log \delta \propto \log M_w$ is observed with a slope of 0.58, in agreement with good solvent condition for a PEG–water system [53]. This suggest that the adsorbed PEG chains on

DMPC/DMPG bilayers exhibit nearly Gaussian statistics, with no strong collapsing, flattening, or stretching due to a high entropic barrier, similar to the case of attractive polymer nanoparticle mixtures [54]. The thickness of the single bilayer was also determined to be ≈6.28 nm based on SAXS results (see Figure 1D), which is close to the hydrodynamic bound polymer layer for ≈20 kDa.

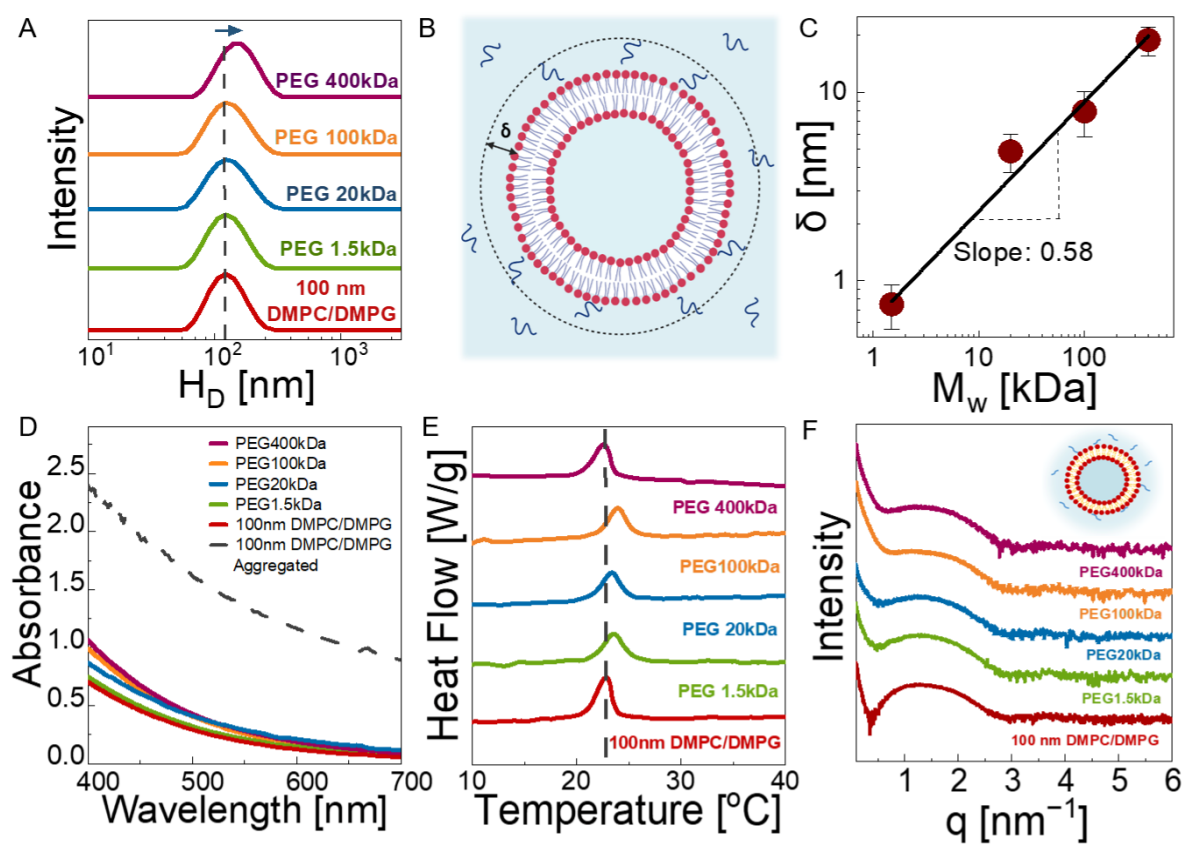

**Figure 2.** (**A**) Hydrodynamic diameter of 2 wt% 100 nm DMPC/DMPG liposomes mixed with 2 wt% PEG solutions with molecular weights of 1.5, 20. 100, and 400 kDa at 35 °C (**B**) Schematic representation of the attachment of PEG chains on lipid bilayer surfaces. δ represents the thickness of the adsorbed layer. (**C**) Molecular weight dependence of δ in PEG–liposome solutions. The data are averaged over three separate estimates; error bars represent one standard deviation. (**D**) UV-vis spectroscopy results of 2 wt% 100 nm DMPC/DMPG liposomes mixed with 2 wt% PEG solutions at 35 °C. The dashed line represents aggregated vesicles after 1 week of preparation. (**E**) DSC results of neat 2 wt% 100 nm DMPC/DMPG liposomes and 2 wt% 100 nm DMPC/DMPG liposomes mixed with 2 wt% PEG solutions at 35 °C. (**F**) SAXS results of neat 2 wt% 100 nm DMPC/DMPG liposomes and mixed with 2 wt% PEG solutions at 35

°C. The inset shows the attachment of PEG chains onto lipid bilayer. All measurements in A, D, E, and F were repeated three times to ensure reproducibility of the data.

The liposome dispersion in concentrated polymer solutions were assessed using UV-vis spectroscopy. Figure 2D shows the absorption spectra of the neat 100 nm liposomes and their composite solution with PEGs with different molar masses (solid lines). Neat liposome solutions result in the lowest wavelength-dependent absorbance typical of spectra from individually dispersed nanoparticles in solution. The absorption increases slightly in the PEG solutions, which is due to the size increase in the liposomes (as suggested by DLS). We ruled out an aggregation scenario as the cause of this increase in intensity by examining the UV-vis spectra (dashed line in Figure 2D) of a neat liposome solution after 1 week of preparation, representing the aggregated state (see Figure S3B for the visuals). We observed a dramatic (about 3-fold) increase in absorbance in the aggregated sample with respect to those dispersed liposomes in polymer solutions.

The SAXS results (Figure 2F) show that 100 nm vesicles remain unilamellar in the presence of PEGs, as no Bragg peak appears. The slope of the curves at low $q$ becomes steeper as the molecular weight of the polymer increases due to excess scattering from PEG macromolecules (see Figure S4 for comparison of the intensity profiles from lipid-free 10 kDa PEG and polymer-free 100 nm liposome solutions). In a previous study by Bandara et al., it was observed that unilamellar DOPC liposomes turned into multilamellar liposomes in the presence of crosslinked PEG chains due to the osmotic pressure applied by polymers into lipid bilayer [55]. However, in the uncrosslinked solution state, we did not observe such a structural change for liposomes, even under the highest osmotic pressure with PEG 1.5 kDa. This shows that not only the osmotic pressure but also the elastic stresses and confinement can play a significant role in the multilamellarity of the liposomes. The unilamellarity of the liposomes in polymer solutions was also evidenced by their thermal transition behavior shown in Figure 2E. No characteristic features of the multilamellar structure (as seen in Figure 1B for solutions of 200 and 400 nm liposomes) was observed, regardless of the PEG chain length, confirming the SAXS results. The change in $T_c$ (≈1 °C) with PEG molar mass is negligibly small. However, Table 1 shows ≈30% decrease in the specific enthalpy for all polymer–liposome solutions with respect to the neat liposome case. A reduction in enthalpy suggests that PEG caused a change in lateral packing of membranes [56], confirming the physical attachment of the PEG chains on the bilayer surfaces.

**Table 1.** Hydrodynamic diameter, polydispersity index (PDI), crystallization temperature ($T_c$), and translational diffusion constant ($D_T$) at 35 °C and 15 °C of neat 20 mg/mL 100 nm DMPC/DMPG liposomes and in PEG solutions with varying molecular weights.

| | $H_D$ (nm) (35 °C) | PDI | $T_c$ (°C) | Specific Enthalpy (J/g) | $D_T$ (m²/s) (35 °C) (x 10$^{-12}$) | $D_T$ (m²/s) (15 °C) (x 10$^{-12}$) |
|---|---|---|---|---|---|---|
| Liposome only | 119.9 ± 1.11 | 0.075 ± 0.010 | 22.84 ± 0.11 | 0.335 ± 0.013 | 5.21 ± 0.018 | 3.73 ± 0.013 |
| In PEG 1.5 kDa | 121.4 ± 0.20 | 0.074 ± 0.009 | 23.55 ± 0.19 | 0.242 ± 0.015 | 5.12 ± 0.020 | 3.61 ± 0.015 |
| In PEG 20 kDa | 129.6 ± 1.10 | 0.077 ± 0.014 | 23.39 ± 0.02 | 0.241 ± 0.017 | 5.03 ± 0.023 | 3.58 ± 0.016 |
| In PEG 100 kDa | 135.7 ± 2.13 | 0.113 ± 0.022 | 23.95 ± 0.32 | 0.234 ± 0.020 | 5.02 ± 0.028 | 3.21 ± 0.020 |
| In PEG 400 kDa | 157.7 ± 3.28 | 0.150 ± 0.025 | 22.58 ± 0.07 | 0.256 ± 0.012 | 4.40 ± 0.019 | 2.72 ± 0.014 |

In summary, the structural analysis verifies that the 100 nm vesicles remain nearly monodisperse and unilamellar in the presence of PEG chains with a molar mass ranging from 1.5 kDa to 400 kDa. The structural uniformity in these complex fluids is essential (although not reported in many previous works) to explicitly understand the role of PEG chains in membrane dynamics.

*3.2. Dynamics*

In this section, we investigate how PEG chains affect the dynamics of 100 nm unilamellar DMPC/DMPG liposomes at different time scales: nanoseconds to milliseconds. We first examined the effect of PEG chains on the translational bulk diffusion of the neat liposomes at 35 °C, when liposomes are in the fluid phase, and at 15 °C, when they are in the gel phase. Then, the viscosity of the solutions is discussed using rheology to understand the effect of particle softness of the liposome particles on the bulk viscosity of the composite system, which exhibits unique thermal stiffening. Finally, the fluorescence lifetime spectroscopy (FLS) results are presented to discuss the microviscosity of the membranes to show how lipid bilayer dynamics correlate with the viscosity of the bulk medium.

### 3.2.1. Bulk Diffusion of Vesicles

Liposomes undergo Brownian motion, resulting in their translational diffusion in solution, as they are colloidal nanoparticles in spherical shapes (i.e., lipid nanoparticles) [57]. Although the structural size analysis was conducted on liposomes by DLS, the translational diffusion coefficient can also be conveniently calculated based on the decay of the autocorrelation functions ($G_2$). In Figure 3A,B, $G_2$ with respect to the delay times are displayed for the liposomes in water and in dilute polymer solutions (0.2%) at 15 °C and 35 °C. In all cases, the relaxation behavior shows a single exponential decay, representing a simple diffusion behavior. Diffusion constant ($D_T$) is calculated by fitting the intensity–intensity autocorrelation function, $G_2(q,\tau)$, which is directly measured by DLS, with a single exponential decay: $G_2(q,t) = A\exp(-2\Gamma\tau)$, where $\Gamma$ is the relaxation rate and related to the diffusion constant by $\Gamma = -D_T q^2$, and q is the wave vector. Figure 3C shows the diffusion coefficients of the polymer-free liposomes; a nearly 1.5-fold decrease was observed as liposomes passed into the gel phase at 15 °C as the membranes turned from a deformable fluid state to a gel state, yielding a rather hard sphere. The estimated diffusion coefficients were found to be $5.21 \times 10^{-12} \pm 0.018$ m$^2$/s and $3.73 \times 10^{-12} \pm 0.013$ m$^2$/s at 35 °C and 15 °C, respectively, and in agreement with those reported in previous studies [58,59].

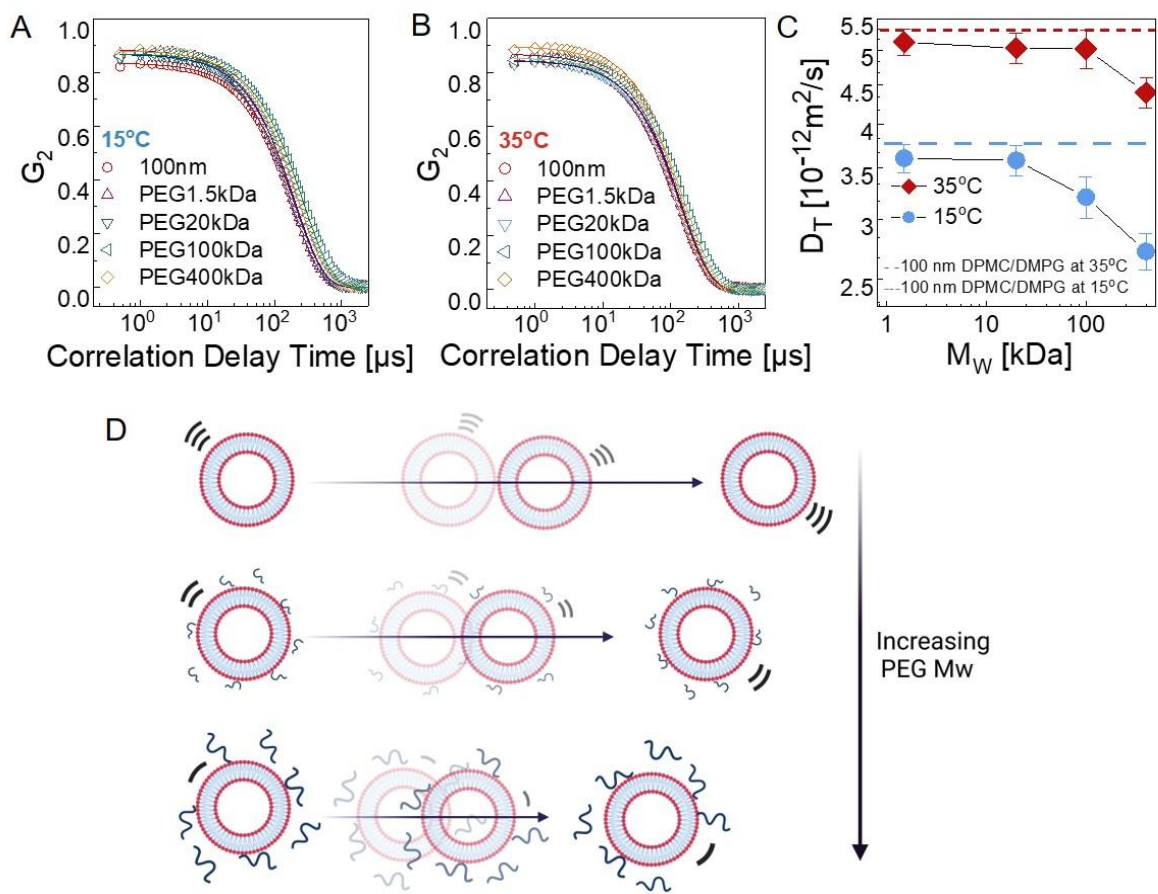

**Figure 3.** Autocorrelation functions of neat 100 nm DMPC/DMPG liposomes and 100 nm DMPC/DMPG liposomes mixed with PEG solutions at (**A**) 15 °C and (**B**) 35 °C. (**C**) PEG molecular-weight-dependent diffusion constants of liposomes at 15 °C and 35 °C. The data are averaged over three separate measurements. (**D**) Schematic representation of the effect of PEG chains on translational liposome diffusion.

The diffusion constant is more affected by the PEG chains when the bilayer is in the gel state. Figure 3C shows the PEG $M_w$ dependence of $D_T$, where significant slowdown starts at $M_w > 20$ kDa at 15 °C, whereas it remains unaffected even up to $M_w \approx 100$ kDa in the fluid state (35 °C) of the bilayer. In the fluid phase, the mesoscale membrane undulations and the polymer's viscous motion occur at comparable time scales; thus, their motions are coupled [13]. Chain diffusion may lead to accelerated liposome motion, leading to the observed trend shown Figure 3C. On the other hand, in the gel state of the bilayers, the liposomes behave more like hollow spheres with solid walls; thus, they act as hard nanoparticles. Their translational diffusion is not facilitated by fast polymer chain relaxation; rather, they are slowed down by the local viscosity increase due to adsorbed polymer. The diffusion constants were obtained in dilute polymer solutions where the bulk concentration of the PEG is small. We verified this by measuring the viscosity of the bulk polymer solutions at the same concentrations studied and observed that the viscosity of the medium remains the same for each case. Thus, the observed slowdown results from the local viscous effect of the physically bound PEG on the liposomes. Although PEGylated liposomes are preferred for drug delivery systems due to increased stability and circulation time by covalent bonds, these results have important practical implications and can be used as a useful tool for rational design of liposome-based drug delivery systems, as the diffusive motion of the lipid vesicles can be effectively controlled by simply changing the length of the attached polymer chains.

3.2.2. Macroscopic Viscosity

Lipid vesicles exhibit several morphologies depending on external parameters, with temperature being the most significant. As the melting temperature of the lipids are passed across, the liposomes are transformed from hard/impermeable spheres with rather rigid walls to soft/permeable spheres with deformable membranes. Such a transition can induce unprecedented effects on the rheology of the lipid nanoparticle solutions [60], and whether it is translated to the complex polymer–liposome solutions must be elucidated.

We first compared the temperature-dependent viscosity of 400 nm, 200 nm, and 100 nm liposome solutions (2 wt% lipids) without polymer to investigate the effect of particle size and lamellarity. Figure 4A displays the relative viscosity of the solutions with respect to water background in the temperature range from 10 to 40 °C (the absolute viscosity values are given in Figure S6A; the viscosity of the deionized water under the same conditions is also provided). Remarkable features are observed, even in these neat liposome solutions. First, the addition of liposomes to water increases the bulk viscosity of the solutions at all temperatures, as expected based on hydrodynamics, and this effect depends strongly on size, lamellarity, and softness of the particles. In the gel phase of the lipids (T << $T_m$), the particles act as solid spheres, and the large vesicles induce greater change in viscosity compared to small vesicles. As the nominal lipid concentration is identical in the samples, the viscosity decreases with size and can be related to the decrease in the effective volume fraction of the particles. At T >> $T_m$, the size effect is enhanced, and the viscosity shift becomes more prominent for the larger liposomes. Because the structure changed from multilamellar to unilamellar with decreasing size (Figure 1D), not only the size but also the lamellarity could play a significant role. This is clearly observed in the viscosity change near $T_m$. For each liposome size, the viscosity rapidly increases as the lipid bilayers transition from gel to fluid state, which is related to the ripple phase of the liposomes [61,62]. The magnitude of the thermal thickening becomes more severe for 400 nm liposomes containing a large fraction of multilamellar structures (see Figure 1D), in which about an 80% increase in relative viscosity is observed compared to less than 10% change for 100 nm unilamellar vesicles. In addition, the transition becomes broader (about seven degrees for 400 nm multilamellar liposomes and two degrees for unilamellar 100 nm vesicles, which is in agreement with the DSC results; see also Figure 1B). Clearly, the structural order within a single bilayer is lost more rapidly compared to the loss of the longer-range order between the stacked bilayers in the multilamellar case. More importantly, by focusing on the unilamellar vesicle solution it is observed that the relative viscosity decreases with temperature in the gel state of the lipids, whereas the trend is reversed in the fluid phase, and a thermal thickening (relative to the neat water) is observed. Above $T_m$, the particles are softer and deformable; thus, the collective lipid motion with low bending elasticity allows for reconfiguration of the structure under deformation and exhibits greater resistances to flow. As the dynamics become faster with temperature, the resistance becomes stronger; thus, the relative viscosity increases with temperature. The results show that the gel-to-fluid transition of the simple lipid bilayers without any other chemical constituents, such as cholesterol, can yield a thermal stiffening of the respective bulk solutions, which can be important to achieve visco-static properties at the cell level, which can contribute to overall homeostasis.

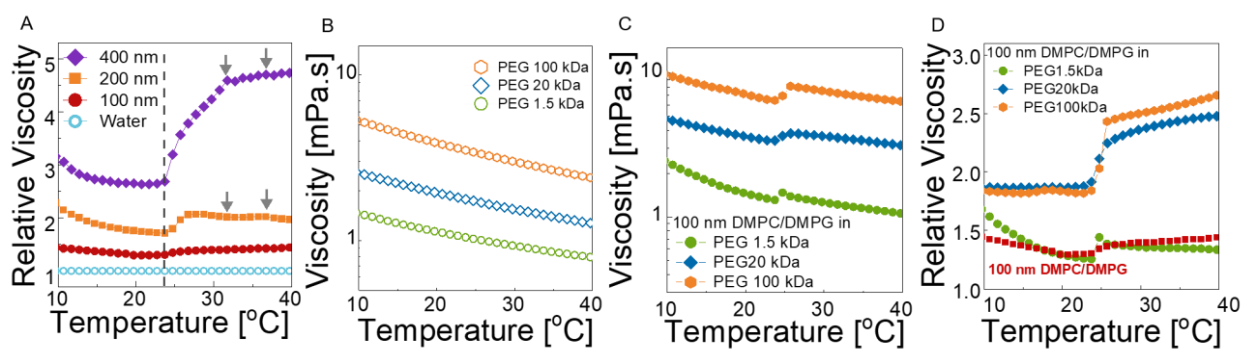

**Figure 4.** (**A**) Relative viscosity of 400 nm, 200 nm, and 100 nm liposomes with respect to water. Vertical dashed line shows the phase transition temperature. Viscosity of (**B**) neat 1.7 wt% PEG solutions with varying $M_w$s (1.5 kDa, 20 kDa, and 100 kDa) and (**C**) composite solutions containing 1.7 wt% lipids and 1.7 wt% polymer in water. (**D**) Relative viscosity of composite solutions containing 1.7 wt% lipids and 1.7 wt% polymer in water with respect to neat PEG solutions. All measurements were repeated three times to ensure reproducibility of the data.

A similar viscosity transition is also observed in the liposome–polymer complexes. Figure 4B shows the viscosity of neat 1.7 wt% PEG solutions for each molecular weight. In the absence of liposomes, a typical polymer solution behavior is apparent, i.e., decreasing viscosity with increasing temperature and with decreasing polymer molecular weight. In the complex solution of 100 nm liposomes and PEG, viscosity shifts to higher values compared to the neat liposome solution and polymer solution viscosities (Figure 4C). Therefore, the liposomes act as reinforcing particles for the neat polymer solutions by causing a nearly 2-fold increase with a dramatic change in the transition regime.

In order to clarify the effect of polymer chain length on the bulk viscosity of liposomes, the relative viscosity of each solution is calculated with respect to the neat PEG solutions (see Figure 4D). In 20 kDa and 100 kDa PEG solutions, the relative viscosity remains nearly unchanged up to the transition temperature, and a strong thermal-thickening regime is observed thereafter. Whereas the slight size increase for the neat liposomes can explain the small increase in viscosity above the melting temperature, strong temperature-dependent thickening in liposome–polymer solutions cannot be explained by the size increase in liposomes only. The magnitude of the viscosity shift at the transition temperature is amplified with increasing PEG chain length (Figure 4D) and increasing concentration of the liposomes (Figure S7), suggesting that the interaction of the polymers bound on liposomes may be important. As the liposomes are individually dispersed in polymer solutions, the average face-to-face distance between the liposomes can be estimated based on the random distribution of spherical particles by $L = d_{DMPC} [(2/\pi\Phi)^{1/3} - 1]$, where L is the face-to-face distance, and $d_{DMPC}$ and $\Phi$ are the diameter and volume fraction of the liposomes, respectively. For a 1.7 wt% nominal lipid concentration and ≈100 nm diameter, the volume fraction of the liposomes (including water enclosed inside) is 9.2% (see Table S3), resulting in an average distance of ≈90 nm. The DLS data (Figure 2A,C) show that the thickness of the bound polymer layer is on the order of the radius of gyration of the PEG chains, suggesting a monolayer coating. Assuming the PEG chains on liposomes adopt Gaussian statistics, the chains occupy $2R_g$ distance from the surface. Thus, for 100 kDa PEG (with $2R_g \approx 23$ nm), the distance between the bound polymers on different particles is about 44 nm. Although this number is an ensemble average over many particles at any instant (or a long time average of a single particle exhibiting Brownian motion), it is too large for bound chains (on different particles) to interact at $T < T_m$ because they are pinned on the hard bilayer surfaces and move together with the liposomes. However, when the bilayer transitions to the fluid state, interaction of the bound chains becomes more likely due to thermally induced bending and thickness fluctuations of the soft membranes, which allows bound polymers to explore longer distances. The interaction of the bound chains can cause a transient network (like entanglement), which results in higher viscosities at $T > T_m$. Additionally, as the polymer–polymer interaction is enhanced with membrane softening, higher temperatures result in enhanced polymer–polymer interaction and thermal thickening. As the chains get shorter, interaction of the bound polymers is less likely; thus, the viscosity shift and thermal transition is less pronounced. On the other hand, in the case of 400 kDa (see SI), $2R_g$ is ≈46 nm, meaning that the bound polymers can overlap, even at $T < T_m$, and the trapped entanglements are released at the melting transition, resulting in large decrease in viscosity as opposed to shorter chains. However, the thermal thickening is still observed at $T > T_m$.

We further investigated the reinforcing behavior of the liposomes below and above the melting temperature of the liposomes in the absence and in the presence of PEG. We used 100 kDa PEG at a fixed polymer concentration of 1.7 wt% and varied the liposome fraction.

We used two models, Einstein and Krieger–Dougherty models, for neat 100 nm liposome solutions and liposomes and PEG 100 kDa solutions at 2.5 wt%, 2 wt%, 1.5 wt%, and 1 wt% liposome concentrations. The viscosity data are presented in Figure S7. The Einstein equation assumes rigid spheres, where the movement of the spheres does not influence each other, and the relative viscosity is given by $\eta_r = 1 + 2.5\phi$, where $\eta_r$ is the relative viscosity, and $\phi$ is the volume fraction of spheres in solution. Krieger–Dougherty is commonly used for the concentrated suspensions [63] with $\eta_r = (1 - \frac{\phi}{\phi_{max}})^{-A\phi_{max}}$. Here, $-A$ and $\phi_{max}$ are 2.7 and 0.71, respectively, for spheres with submicron sizes [64]. In our case, the relative viscosity is measured with rheology, and the volume fractions are calculated based on the nominal lipid concentration, hydrodynamic size, and area per lipid information. We analyzed the data at 35 °C and 15 °C, well above and well below the melting temperature, respectively. The total number of liposomes is calculated from the total surface area and area per lipid of DMPC membranes in fluid and gel phases. The repeat distance of bilayers, including the water phase at 15 °C, is obtained from SAXS as 6.82 nm (see Figure S8). The details of the volume fraction calculation are given in Supporting Information. The symbols in Figure 5A,B show the calculated volume fractions and their corresponding relative viscosity values. The lines represent the predictions from the models. Clearly, the Einstein model fails to explain the observed trend, whereas the Krieger–Dougherty model reasonably accounts for the neat liposome data in the fluid and gel states. The viscosity shift observed for the 100 nm liposome solutions is therefore mainly related to the hydrodynamic size increase in the liposomes, which effectively increases the particle volume fraction. In the presence of PEGs, the Krieger–

Dougherty model still adequately explains the observed behavior at 15 °C (except only for the solution with the largest liposome volume fraction, which diverges from the prediction), whereas at 35 °C, the model significantly overestimates the effective volume fraction of the liposomes under all nominal lipid concentrations, suggesting that the size increase in the liposomes in the fluid state alone is not sufficient to explain the viscosity shift; the particles change from hard nanoparticles to soft nanoparticles [65]. As the model still predicts the viscosity trend in the fluid phase of the neat liposome solutions, the soft particle behavior is mainly due to the presence of PEG chains that are dynamically coupled with the flexible membrane. The 'smart' temperature-responsive behavior of liposome–polymer solutions driven by particle softness is intriguing and demands more detailed theoretical and experimental investigations.

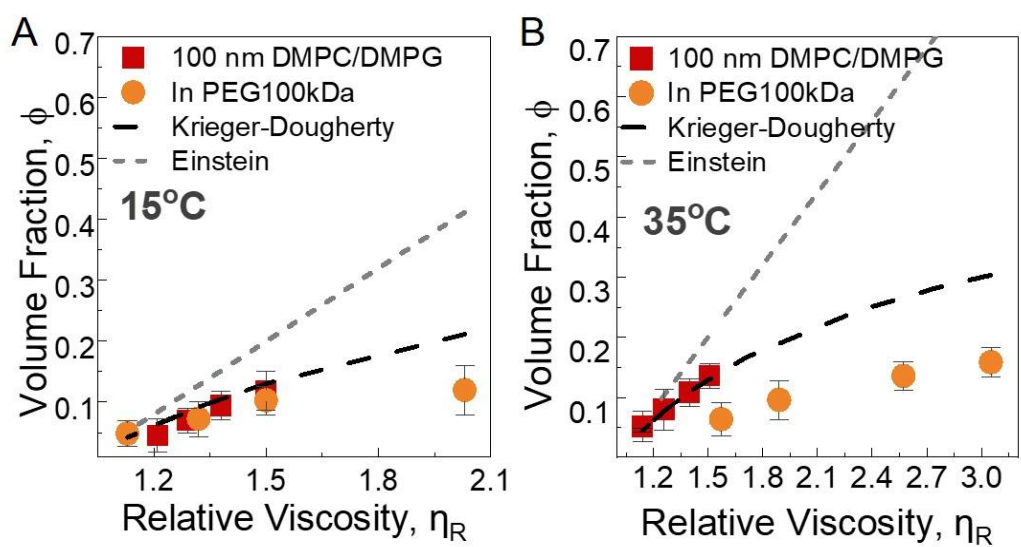

**Figure 5.** Volume fractions versus relative viscosity of the neat 100 nm liposome solutions and liposomes in 1.7 wt% PEG (100 kDa) solutions at four lipid concentrations (2.5 wt%, 2 wt%, 1.5 wt%, and 1 wt%). The curves show the predictions of Einstein and Krieger–Dougherty viscosity models at (**A**) 35 °C and (**B**) 15 °C. The data are averaged over three separate measurements, and the error bars represent one standard deviation.

3.2.3. Microscopic Viscosity in Bilayers

Finally, we investigated how the presence of PEG chains influences the microviscosity of the lipid bilayers using time-resolved fluorescence spectroscopy at 35 °C and 15 °C by staining lipid bilayers with Oil Red O. Estimation of the microviscosity of the lipid membranes is a well-studied technique using fluorophores [66–68]. Oil Red O is a hydrophobic fluorescent probe that is located within the membrane and widely used to monitor cell membranes for biological applications [69]. The lifetime of the probe depends on the viscosity of its local microenvironment (as expressed by the Förster–Hoffman theory [70]) according to the following equation: $\ln(\tau) = A + B \ln(\eta)$, where $\tau$ is the fluorescent lifetime (FL) of the Oil Red O, $\eta$ is the viscosity, and A and B are the intercept and slope of the line, respectively. The purpose of constructing a calibration curve is to determine the changes in the lifetime of the probe in the environments with known viscosities and to obtain the constants of the Förster–Hoffman equation. Methanol–glycerol solutions with a wide range of viscosity values were used because Oil Red O is able to dissolve in both methanol and glycerol, which is not possible with a water–glycerol mixture due to the insolubility of Oil Red O, and it does not form aggregations. There are many examples in previous works studying microviscosity of the lipid bilayer using hydrophobic fluorophores with methanol–glycerol or ethanol–glycerol solutions [67,71]. Solution viscosities were measured with a shear rheometer at 15 °C and 35 °C (see Figure S9 for raw viscosity-shear rate data), and the values were used in the Newtonian flow regime. For each of the solutions, the lifetimes of Oil Red O were obtained and plotted against the viscosity (see Figure 6A,B). The linear dependence at both temperatures, confirms the validity of the Förster Hoffman equation for this system. The constants of the equation were obtained as 0.0145, and 9.9327 for A, and 10.206, and 0.0457 for B at 15 °C and 35 °C, respectively. We then performed fluorescent lifetime measurements on the neat 100 nm liposome solutions and on 100 nm liposome solutions in the presence of PEGs with varying $M_w$ values at 15 °C and 35 °C (see Figure 6C,D.) The excited-state lifetime of the probe is perturbed by local fields and neighboring dipoles [72,73]. Therefore, spatial identification of the measuring point and the fluorescence probe is crucial to monitor the targeted chemical and biological processes [68,74]. The hydrophobic nature of our probe and the Oil Red O maintains their localization within the membrane, which enables the estimation of microviscosity of the lipid bilayers. The individual lifetime values are presented in Tables S4 and S5, and the average lifetimes calculated based on the exponential decay and the estimated microviscosities within the lipid bilayers are presented in Table 2.

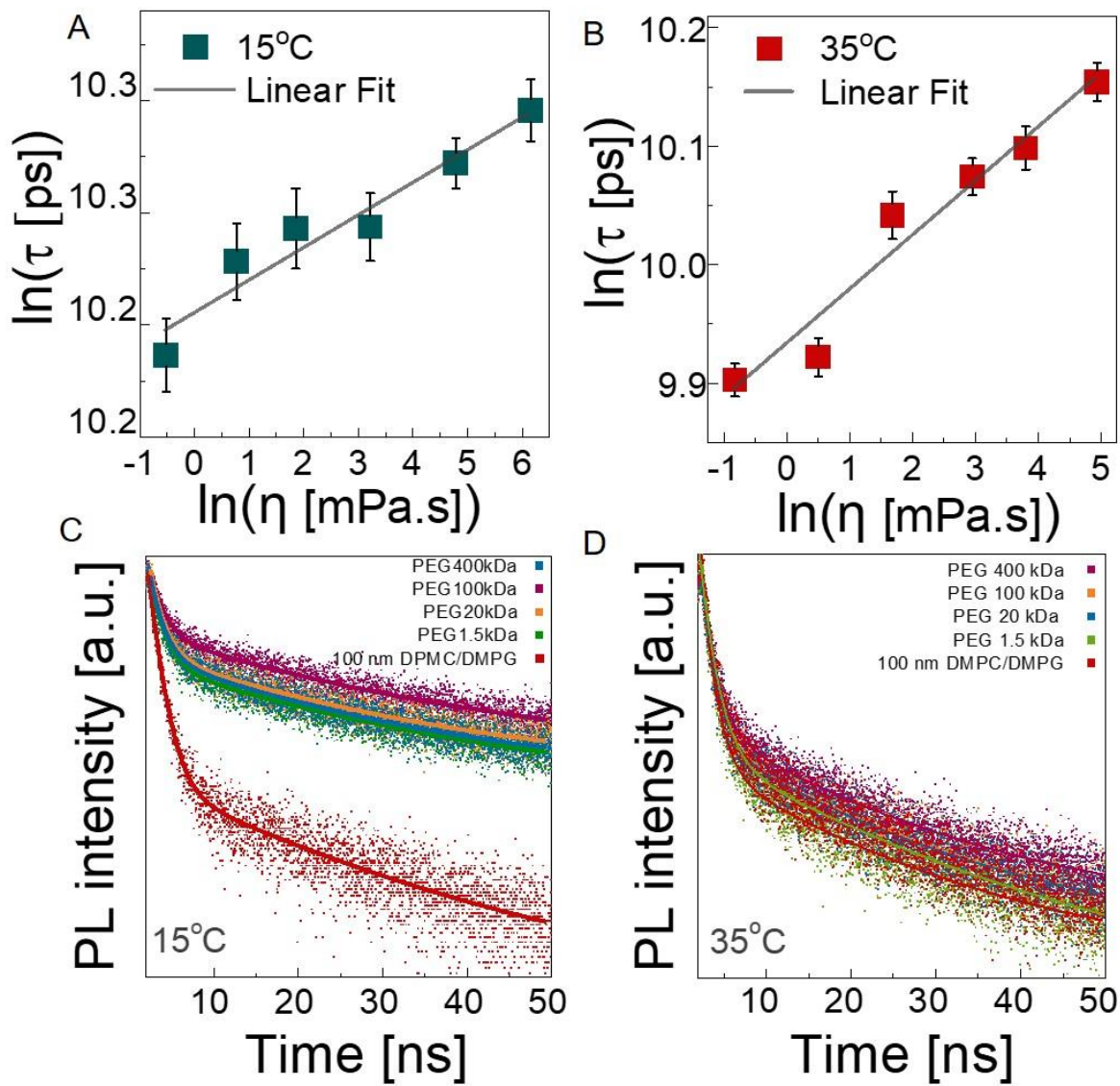

**Figure 6.** Average fluorescence lifetime of Oil Red O in methanol–glycerol mixtures of various compositions plotted against viscosity at (**A**) 15 °C and (**B**) 35 °C. Error bars represent standard deviation; measurements were repeated three times. Time-resolved fluorescence decays of Oil Red O inside lipid bilayer in the presence of different PEG $M_w$ values at (**C**) 15 °C and (**D**) 35 °C.

**Table 2.** Lifetime of Oil Red O and microviscosity of lipid membranes in 100 nm liposome solutions in the absence and presence of PEG with various $M_w$ values.

| | τavg (ns) 35 °C | Microviscosity (cP) 35 °C | Bulk Viscosity(cP) 35 °C | τavg (ns) 15 °C | Microviscosity (cP) 15 °C | Bulk Viscosity(cP) 15 °C |
|---|---|---|---|---|---|---|
| 100 nm liposomes | 24.01 ± 0.032 | 28.75 ± 0.85 | 1.414 | 28.91 ± 0.038 | 142.82 ± 13.54 | 1.546 |
| in PEG 1.5 kDa | 20.61 ± 0.013 | 1.02 ± 0.01 | 1.156 | 27.38 ± 0.033 | 3.36 ± 0.29 | 1.827 |
| in PEG 20 kDa | 21.43 ± 0.022 | 2.39 ± 0.05 | 3.406 | 27.59 ± 0.036 | 5.69 ± 0.53 | 4.163 |
| in PEG 100 kDa | 22.72 ± 0.019 | 8.59 ± 0.15 | 6.874 | 28.42 ± 0.029 | 43.93 ± 3.2 | 8.247 |
| in PEG 400 kDa | 23.91 ± 0.028 | 26.24 ± 0.68 | 23.94 | 28.74 ± 0.031 | 95.09 ± 7.34 | 53.35 |

In the absence of PEG, the lifetime of the probe decreases from ≈28.91 ns in the gel state to ≈24.01 ns in the fluid state, corresponding to microviscosities of 142.82 cP and 28.75 cP at 15 °C and 35 °C, respectively. These values are in agreement with previously reported data for DMPC liposomes [40,75]. The control FLS experiments with Oil Red O mixed with PEG solution in water (see Figure S10) revealed that average fluorescence lifetime of liposome Oil Red O conjugation is higher than the average fluorescence lifetimes for these control media due to increased interaction pathways.

Addition of PEG into liposome solution causes chain-length-dependent changes in lifetime. Figure 7 compares the microviscosity of liposomes for different PEG $M_w$ values at 15 °C and 35 °C. The horizontal lines indicate the values obtained from polymer-free liposome solutions. In a solution of 400 kDa PEG, the microviscosity is close to that for the neat liposome solution, and it monotonically decreases by about two orders of magnitude with decreasing chain length. In the fluid state, the overall trend is similar but with different magnitudes. Clearly, the bilayer microenvironment is strongly influenced by the surrounding PEG chains and the viscosity of the medium, which is finely tunable with the length of the polymer chains and their concentrations. As the chains get shorter, the relaxation becomes faster, and the dynamical coupling between the membrane and the polymer segments is more pronounced, which may facilitate the thickness fluctuations and/or membrane undulations. Recently, Nagao et al. [76]

measured the acyl tail dynamics of DMPC at broad time scales at the relevant length scale using high-resolution neutron spectroscopy and X-ray scattering techniques and correlated it with membrane viscosity. Similarly, using deuterated PEG and tail-deuterated DMPC in $D_2O$, the structural dynamics of the acyl tails and therefore the membrane viscosity, can be more directly obtained in these complex fluids to further elucidate the underlying molecular mechanisms, which we will focus on in a future study.

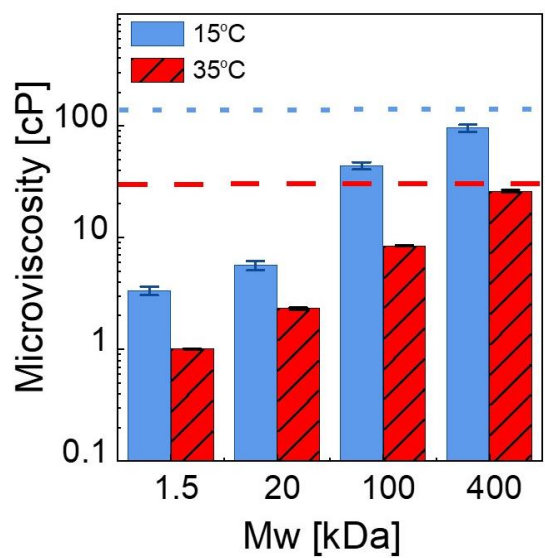

**Figure 7.** Comparison of microviscosity of the lipid bilayers in 2 wt% PEG solutions with different molecular weights at 15 °C and 35 °C. Short dotted lines and dashed lines indicate the microviscosity of the neat liposomes at 15 °C and 35 °C, respectively. The data are averaged over three separate measurements, and the error bars represent one standard deviation.

We therefore conclude that the size increase observed in DLS for low PEG concentrations is due to a bound PEG layer, the thickness of scales with the expected Mw dependence of PEG. Additionally, if the osmotic pressure was effective at our concentration, the size should have decreased, as observed at higher concentrations in DLS.

It is known that the addition of PEG into a liposome solution can create an osmotic pressure and induce dehydration or aggregation of vesicles, which may result in enhanced lateral packing in the membrane [77,78]. Although osmotic pressure is effective at relatively higher concentrations, we assessed its possible influences on our results. We performed DLS experiments with liposomes in 1.5 kDa and 20 kDa PEG solutions (where the osmotic pressure would be greater) at varying PEG concentrations (from 0.2 wt% to 10 wt%) to assess the effect of osmotic pressure on size (see Figure S11). We observed that the addition of polymers at low concentrations increases the hydrodynamic diameter of the particles due to the attached PEG layer, and the size remains unchanged up to ≈3%. In the case of 20 kDa PEG, the size shift is more pronounced compared to 1.5 kDa PEG, a which results in smaller bound layer. At concentrations above ≈3%, the particle size decreases appreciably due to dehydration of liposomes by the osmotic pressure. At even higher concentrations, a significant size increase (possibly due to fusion or aggregation) is observed. These concentrations are far from the concentration used in our work (1.7 wt%). Note also that the sizes obtained from DLS were estimated based on the Stokes–Einstein relation using the measured viscosity of the polymer solutions; therefore, the influence of solvent viscosity on size estimation is eliminated. Therefore, we conclude that the size increase observed in DLS for low PEG concentrations is due to the bound PEG layer, the thickness of which increases with the expected size dependence of PEG $M_w$ (Figure 2C). The presence of PEG layer on liposomes was also confirmed by measuring their zeta potentials as a function of PEG concentration (see Table S6). The PEG layer on liposomes causes a shift in the hydrodynamic slip layer; the zeta potential increases from ≈−9 mV to ≈0 with the addition of PEG.

In order to evaluate the osmotic pressure effect on bulk thermal thickening, we prepared an additional liposome–PEG solution by hydrating the lipids with 20 kDa PEG solutions to form liposomes containing PEG both inside and outside of the vesicles. The solutions were then extruded using the same protocol to obtain liposomes with identical size and polydispersity (confirmed with DLS). Figure S12 shows no difference in the transition behavior of liposomes when compared to the data presented in Figure 2C (for empty liposomes at the same PEG concentration). Therefore, the observed viscosity shifts are not related to the osmotic pressure variation but due to the hard-to-soft particle transition, as discussed above. Regarding the microviscosity, a recent study by Sahu et al. suggests that the microviscosity of unilamellar DMPC vesicles increases with the addition of PEG-400 (400 Da) and PEG-6000 (6 kDa) [26]. The authors attributed this to the increase in the lateral packing density in the bilayers in the PEG environment. The concentration of the PEGs used in their study was 25 wt%, which is sufficiently high to create osmotic pressure and cause aggregation/fusion. This is unlike what we observed in our study, where the osmotic pressure was not as effective due to low concentration of PEGs. Instead, the addition of PEG chains caused a dynamic coupling and accelerated lipid dynamics. Finally, we note that most liposome-based drug carriers utilize PEGylated lipids, where PEG is covalently attached to lipid to enhance colloidal stability and blood circulation of the nanoparticles. In such systems, a persistent polymer layer is developed on liposome surfaces, and the interaction of graft PEG chains with membrane lipids differs relative to the physisorbed PEG, as the conformation of end-tethered polymers can be significantly altered depending on the chain length and graft density. This study provides fundamental insights for future studies concerning many other aspects of polymer–membrane interactions.

**4. Conclusions**

In summary, using aqueous solutions of poly(ethylene glycol) (PEG) with a wide range of molecular weights (from 1.5 kDa to 400 kDa) and nearly monodisperse 100 nm unilamellar vesicles of DMPC/DMPG liposomes, we studied the effect of polymeric media on the structure and dynamics of lipid membranes from the bulk scale to the nanoscale. Although the addition of PEG chains did not cause a notable structural change, as confirmed by SAXS and DSC curves, they attached to liposome surfaces, resulting in an adsorbed polymer layer on bilayers. Translational diffusion constants of liposomes revealed that an adsorbed PEG layer is more affected by mesoscale membrane undulations in the fluid phase, where viscous motion of liposomes and polymers is coupled at comparable time scales. At the macroscopic level, liposomes act as reinforcing particles for polymer solutions with a remarkable increase in viscosity at the gel-to-fluid transition temperature and thermal thickening thereafter. The microviscosity of the lipid bilayers is directly affected by surrounding PEG chains and therefore by the relaxation of the whole chain, resulting an enhancement of the membrane dynamics in the case of short chains due to the fast relaxation time of the shorter chains, which enables dynamic coupling of membrane and polymer motion. Overall, we showed that nanoscale membrane dynamics, diffusive motion, and rheological behavior of liposomes can be effectively tuned by changing the polymer chain length, concentration, and lamellarity.

**Supplementary Materials:** The following supporting information is available.

**Author Contributions:** Conceptualization, E.S.; data curation, S.K. (Selcan Karaz) and E.S.; formal analysis, S.K. (Selcan Karaz), M.H., and E.S.; funding acquisition, E.S.; investigation, S.K. (Selcan Karaz), G.A., A.O., S.N. and E.S.; methodology, S.K. (Selcan Karaz), M.H. and G.A.; project administration, E.S.; resources, S.N. and E.S.; supervision, S.K. (Seda Kizilel) and E.S.; validation, S.K. (Selcan Karaz) and E.S.; visualization, S.K. (Selcan Karaz); writing—original draft, S.K. (Selcan Karaz), M.H. and E.S.; writing—review and editing, S.K. (Seda Kizilel) and E.S. All authors have read and agreed to the published version of the manuscript.

**Funding:** This work was supported through the Marie Skłodowska-Curie Actions (MSCA) Widening Fellowship (grant no: 101003358) under the Horizon 2020 Program of the European Commission.

**Data Availability Statement:** The data presented in this study are available on request from the corresponding author.

**Acknowledgments:** The authors acknowledge Central Research Infrastructure Directorate at Koç University for the use of SAXS services and Koç University Surface Science and Technology Center (KUYTAM) for DLS characterization. The authors thank Michihiro Nagao (NIST NCNR) for fruitful discussion.

**Correspondence**: esenses@ku.edu.tr (E.S.), skizilel@ku.edu.tr (S.K)

**Conflicts of Interest:** The authors declare no conflict of interest.